\documentclass[aps,prd,onecolumn,groupedaddress,showpacs,nofootinbib,amssymb]{revtex4}
\usepackage[dvips]{graphicx}
\usepackage{amssymb}
\usepackage{amsmath}
\usepackage{graphicx,,color}
\usepackage{amsfonts}
\usepackage{bm}
\usepackage{cancel}
\usepackage{comment}
\usepackage[dvipsnames]{xcolor}
\usepackage{makecell}

\allowdisplaybreaks[4]

\begin{document}

\tolerance=5000

\title{General Scalar Field Inflation ACT Attractors: Utilizing the $n_s(r)$ relation }
\author{V.K. Oikonomou$^{1,2}$}\email{voikonomou@gapps.auth.gr;v.k.oikonomou1979@gmail.com}
\affiliation{$^{1)}$Department of Physics, Aristotle University of
Thessaloniki, Thessaloniki 54124, Greece\\
$^{2)}$ Center for Theoretical Physics, Khazar University, 41
Mehseti Str., Baku, AZ-1096, Azerbaijan}

\begin{abstract}
The ACT data have severely constrained the single scalar field
models. Known models of inflation, like the Starobinsky model, the
Higgs model and the $a$-attractors are at least $2\sigma$ off the
ACT data. In this work we aim to provide a top-to-bottom approach
in single scalar field inflationary cosmology compatible with the
ACT data. Specifically, inspired by the fact that the Starobinsky
model, the Higgs model and the $a$-attractors, all being plateau
potentials, result to the same attractor relation between the
spectral index of scalar perturbations and the tensor-to-scalar
ratio, which is of the form $n_s(r)=1-\alpha r^{1/2}$, in this
work we seek for attractors of the form $n_s(r)=f(r)$ that may
lead to ACT-compatible inflation. Specifically, we fix the
function $f(r)$ to have a specific desirable form and then solve
the differential equation $n_s(r)=f(r)$ to find the potential
which results to the relation $n_s(r)=f(r)$. We discovered
analytically three classes of potentials which are variants of the
general form $n_s(r)=\gamma \pm \beta r \pm r^{1/2}$ and all these
models are found to be compatible with the ACT data.
\end{abstract}

\maketitle

\section{Introduction}

Inflation \cite{inflation1,inflation2,inflation3,inflation4} in
its various forms is to date the most prominent theoretical
description we have for the primordial era of the Universe.
Inflation occurred at the verge of the quantum gravity era, and it
is believed that the Universe was classical and four dimensional
when it occurred. The inflationary regime of our Universe was in
the focus of the cosmic microwave background (CMB) radiation
experiments the last 30 years and still is, since we have no
smoking gun signal for it yet. However, having a nearly $4\sigma$
red tilted scalar spectral index, a spectrum of nearly Gaussian
perturbations and a flat Universe already proves the quantum
origin of the Universe. Nevertheless the detection of either a
B-mode in the CMB spectrum or a stochastic gravitational wave
radiation background will prove the existence of an inflationary
regime. The former will be intensively sought in the near future
CMB stage 4 experiments like the Simons observatory
\cite{SimonsObservatory:2019qwx} and also by the LiteBird
collaboration \cite{LiteBIRD:2022cnt}, while the stochastic
gravitational wave background will be probed by the future
gravitational wave experiments
\cite{Hild:2010id,Baker:2019nia,Smith:2019wny,Crowder:2005nr,Smith:2016jqs,Seto:2001qf,Kawamura:2020pcg,Bull:2018lat,LISACosmologyWorkingGroup:2022jok}.
The existence of a stochastic gravitational wave background was
confirmed by the Pulsar Timing Array experiments like NANOGrav
\cite{NANOGrav:2023gor} but the signal cannot be explained by
inflation solely  \cite{Vagnozzi:2023lwo,Oikonomou:2023qfz}. The
mystery in the inflationary theories is which theory actually
describes it. There are various proposals for describing
inflation, such as modified gravity
\cite{reviews1,reviews2,reviews3,reviews4}, and there is room for
modifications of general relativity, since the DESI data indicated
phenomena that can be consistently described only by modifying
general relativity, such as the phantom crossing
\cite{Lee:2025pzo, Ozulker:2025ehg, Kessler:2025kju,
Nojiri:2025low, Vagnozzi:2019ezj}. Nevertheless the standard
single scalar field description is the most used description of
inflation. Recently in 2025, the released ACT data shook the
ground of inflationary physics, since the spectral index of the
scalar perturbations was reported to be constrained as follows
\cite{ACT:2025fju, ACT:2025tim},
\begin{align}
\label{act} n_{s}=0.9743 \pm 0.0034\, .
\end{align}
Combined with the updated Planck/BICEP constraints on the
tensor-to-scalar ratio \cite{BICEP:2021xfz},
\begin{align}
\label{planck} r<0.036\, ,
\end{align}
the ACT constraints have rendered many inflationary models to be
non-viable, in one night. Since the ACT data were released, many
studies have appeared in the literature aiming to reconcile
various inflationary models with the ACT data, see for example
Refs.
\cite{Kallosh:2025rni,Gao:2025onc,Liu:2025qca,Yogesh:2025wak,Yi:2025dms,Peng:2025bws,Yin:2025rrs,Byrnes:2025kit,
Wolf:2025ecy,Aoki:2025wld,Gao:2025viy,Zahoor:2025nuq,Ferreira:2025lrd,Mohammadi:2025gbu,Choudhury:2025vso,
Odintsov:2025wai,Q:2025ycf,Zhu:2025twm,Kouniatalis:2025orn,Hai:2025wvs,Dioguardi:2025vci,Yuennan:2025kde,
Kuralkar:2025zxr,Kuralkar:2025hoz,Modak:2025bjv,Oikonomou:2025xms,Odintsov:2025jky,
Aoki:2025ywt,Ahghari:2025hfy,McDonough:2025lzo,Chakraborty:2025wqn,NooriGashti:2025gug,Yuennan:2025mlg,
Deb:2025gtk,Afshar:2025ndm,Ellis:2025zrf,Yuennan:2025tyx,Wang:2025cpp,Qiu:2025uot,Wang:2025dbj,Asaka:2015vza,Oikonomou:2025htz}.
Note however, that some caution is needed in interpreting the ACT
data \cite{Ferreira:2025lrd}.

In this work we will focus on the following question, how can we
construct from scratch ACT-compatible inflationary models in a
model agnostic way. Here, model agnostic is used in a sense that
one does not start from the scalar potential, building the
inflationary phenomenology in a bottom-up approach, on the
contrary we will use a top-bottom approach. Specifically, we shall
seek for attractor properties in general single canonical scalar
field theories, using the functional relation of the scalar
spectral index as a function of the tensor-to-scalar ratio, namely
$n_s(r)=f(r)$. We are motivated by the fact that in order for the
theory to be compatible with the ACT data, one needs to have
considerable control over the scalar spectral index. So by
defining the functional form of the spectral index $n_s(r)=f(r)$
one is lead to the scalar potential by solving the differential
equation $n_s(r)=f(r)$. The theories that give a simple $n_s(r)$
relation are plateau inflation models, like the Starobinsky model
\cite{Starobinsky:1980te}, Higgs inflation \cite{Bezrukov:2007ep}
and $a$-attractor models
\cite{Kallosh:2015lwa,Kallosh:2013maa,Kallosh:2013daa,Kallosh:2014rga}
and all give $n_s(r)=1-\alpha r^{1/2}$. The differential equation
$n_s(r)=f(r)$ can be solved analytically for simple choices of the
function $f(r)$, so in this work we present three distinct
attractor models of the form $n_s(r)=f(r)$ which result to an ACT
and BICEP/Planck compatible phenomenology. The inflationary
phenomenology of each model is analyzed in detail. Of course there
might be extensions of this work, but we aimed to point out the
perspective of having inflationary attractors by utilizing the
relation $n_s(r)=f(r)$.

\section{Known Single Scalar Field Attractors and the $n_s(r)$ Relation}

Most known inflationary models, which are popular in the field,
are $a$-attractors
\cite{Kallosh:2015lwa,Kallosh:2013maa,Kallosh:2013daa,Kallosh:2014rga},
Higgs inflation \cite{Bezrukov:2007ep}, Starobinsky inflation
\cite{Starobinsky:1980te} and similar. All these models lead to an
attractor behavior of the observational indices, which have the
following form,
\[
n_s-1 \simeq -\frac{2}{N},\qquad r \simeq \frac{12}{N^2}, .
\]
All these potentials have a specific shape  which is an
exponentially approached plateau, a feature which explains why
Higgs (non-minimally coupled), Starobinsky, $a$-attractors and
many similar models fall into that class of inflationary models.
For the moment, and for simplicity we will use Planck units, with
\(\kappa=1\), where \(\kappa^2=1/M_{p}^2\) and $M_p$ is the
reduced Planck mass. We will return to natural units when we
consider the ACT attractors later on.

All the models above share the same large-field behavior, that is,
the potential tends to a positive constant \(V_0\) and approaches
it exponentially,
\begin{equation}\label{eq:plateau}
V(\phi)\simeq V_0\Big(1 - A e^{-\lambda\phi} +
\cdots\Big)\qquad(\phi\to+\infty),
\end{equation}
with constants \(A>0,\ \lambda>0\). Now, why is this shape
universal? Non-minimal couplings or pole-like kinetic terms,
actually map the original variables into a canonical scalar field
with an exponential stretching, so the large-\(\phi\) expansion at
leading order, has the universal form \eqref{eq:plateau}. So let
us discover what are the consequences for the slow-roll parameters
and subsequently for the observational indices. We start from the
slightly more precise plateau potential form,
\[
V(\phi)\simeq V_0\big(1 - A e^{-\lambda\phi}\big)^2
\quad\text{or}\quad V_0\big(1 - A e^{-\lambda\phi}+\cdots\big)\, ,
\]
so for large \(\phi\) values, we keep the leading exponential
term,
\[
V'(\phi)\simeq V_0\,A\lambda\,e^{-\lambda\phi},\qquad
V''(\phi)\simeq -V_0\,A\lambda^2\,e^{-\lambda\phi}.
\]
The slow-roll parameters in scalar field inflation are,
\[
\epsilon \equiv \frac{1}{2}\Big(\frac{V'}{V}\Big)^2,\qquad \eta
\equiv \frac{V''}{V}\, ,
\]
hence, at leading order we get,
\begin{align*}
\epsilon &\simeq \tfrac{1}{2}(A\lambda e^{-\lambda\phi})^2 \propto e^{-2\lambda\phi},\\
\eta &\simeq -A\lambda^2 e^{-\lambda\phi} \propto
e^{-\lambda\phi}.
\end{align*}
Note that \(\eta\) decays like \(e^{-\lambda\phi}\) while
\(\epsilon\) decays like \(e^{-2\lambda\phi}\). Hence, in the
plateau regime \(\lvert\eta\rvert\gg\epsilon\), both are small,
and in fact,
\[
\eta \propto -\sqrt{\epsilon}\quad(\text{up to model-dependent
constants}),
\]
because \(\sqrt{\epsilon}\sim A\lambda e^{-\lambda\phi}\) while
\(\eta\sim -A\lambda^2 e^{-\lambda\phi}\). Now, let us calculate
the number of e-folds in slow-roll approximation,
\[
N(\phi)\simeq \int_{\phi_{\rm
end}}^{\phi}\frac{V}{V'}\,d\tilde\phi \;\simeq\; \int
\frac{1}{A\lambda}e^{\lambda\tilde\phi}\,d\tilde\phi
\;\Rightarrow\; e^{\lambda\phi}\propto N\, ,
\]
hence \(e^{-\lambda\phi}\propto 1/N\), and therefore we have,
\[
\epsilon \propto \frac{1}{N^2},\qquad \eta\propto -\frac{1}{N}.
\]
More precisely, one finds
\[
\epsilon \simeq \frac{1}{2\lambda^2 N^2},\qquad \eta \simeq
-\frac{1}{N} + \mathcal{O}\!\big(\tfrac{1}{N^2}\big).
\]
So upon using the leading order slow-roll expressions for the
observables,
\[
n_s - 1 \simeq -6\epsilon + 2\eta,\qquad r\simeq 16\epsilon.
\]
we get
\[
r \simeq \frac{8}{\lambda^2}\frac{1}{N^2},\qquad n_s - 1 \simeq
2\Big(-\frac{1}{N}\Big) + \mathcal{O}\!\big(\tfrac{1}{N^2}\big) =
-\frac{2}{N} + \mathcal{O}\!\big(\tfrac{1}{N^2}\big),
\]
hence the universal leading order predictions are,
\[
\boxed{\,n_s-1 \simeq -\frac{2}{N},\qquad r \simeq
\frac{8}{\lambda^2 N^2}\,.\,}
\]
If \(\lambda=\sqrt{2/3}\) (Starobinsky/Higgs), then
\[
r\simeq \frac{8}{(2/3)N^2}=\frac{12}{N^2},
\]
and the familiar pair
\((n_s,r)\simeq\big(1-\tfrac{2}{N},\,\tfrac{12}{N^2}\big)\)
follows. For \(\alpha\)-attractors \(\lambda^2=2/(3\alpha)\),
hence \(r\simeq 12\alpha/N^2\). Now, let us find the $n_s(r)$
relation for the plateau potentials under consideration. From the
exponential scaling one finds an approximate algebraic relation
between \(\eta\) and \(\epsilon\). Since both scale with the same
exponential factor up to powers,
\[
\eta \simeq -(\sqrt{2}\,\lambda)\,\sqrt{\epsilon} +
\mathcal{O}(\epsilon),
\] using \(r=16\epsilon\), the empirical relation
\(n_s-1\propto -r^{1/2}\) can be written (up to constants) as
\[
2\eta - 6\epsilon \simeq -C\sqrt{\epsilon},
\]
so
\[
\eta \simeq 3\epsilon - \tfrac{C}{2}\sqrt{\epsilon}.
\]
The precise constant \(C\) depends on \(\lambda\) and therefore on
the model. Hence, for plateau potentials, one has the general
relation,
\begin{equation}\label{plateau}
n_s(r)=1-\frac{1}{\sqrt{\alpha}}r^{1/2}\, ,
\end{equation}
where $\alpha$ is a model dependent parameter. This general
attractor behavior in Eq. (\ref{plateau}) is always obtained by
potentials that approach a constant exponentially,
\[
V(\phi)\simeq V_0\big[1 - A e^{-\lambda\phi} + \cdots\big],
\]
which produce the universal attractor predictions,
\[
n_s-1\simeq -\frac{2}{N},\qquad r\simeq \frac{8}{\lambda^2 N^2}.
\]
This is because \(V''/V\sim e^{-\lambda\phi}\) while
\((V'/V)^2\sim e^{-2\lambda\phi}\), the \(\eta\) term dominates
\(n_s-1\) and scales like \(-1/N\). To leading order in the
plateau regime we have,
  \[
  \eta \simeq -(\text{const})\sqrt{\epsilon},\qquad
  \epsilon \sim \frac{1}{2\lambda^2 N^2}.
  \]
For Starobinsky/Higgs inflation (\(\lambda=\sqrt{2/3}\)) this
gives \(r\simeq 12/N^2\).

Now let us reverse the logic, and instead of seeking and studying
the phenomenology of distinct potentials, let us use specific
universal relations for $n_s(r)$, like the one in Eq.
(\ref{plateau}), and find the the potentials that lead to these
relations. Then we can examine the phenomenology of the resulting
potentials in detail. In this way we will have full control on the
spectral index of the scalar perturbations, which we can easily
engineer to be compatible with the ACT data, at least for some of
the potentials.

\subsection{Methodology for General $n_s(r)$ Attractors}

The starting point will be the canonical scalar field action,
\begin{equation}
    \centering\label{act}
    S = \int d^4 x \sqrt{- g}\left(\frac{R}{2 \kappa^2} - \frac{1}{2}g^{\mu \nu} \partial_\mu \phi \partial_\nu \phi - V(\phi) \right),
\end{equation}
We shall assume that the background metric is a
Friedmann-Robertson-Walker (FRW) metric, with a flat spatial part,
\begin{align}
\label{metricflrw} ds^2 = - dt^2 + a(t)^2 \sum_{i=1,2,3}
\left(dx^i\right)^2\, .
\end{align}
The scalar field equation of motion is obtained  by varying the
action with respect to the scalar field,
\begin{equation}
    \centering\label{eomphi}
    \ddot{\phi} + 3 H \dot{\phi} + V' = 0\, .
\end{equation}
Also by varying the action (\ref{act}) with respect to the metric
tensor, we get,
\begin{equation}
    \centering\label{eom}
    \frac{1}{\kappa^2}\left(R_{\mu \nu} - \frac{1}{2}R g_{\mu \nu}\right) = \partial_\mu \phi \partial_\nu \phi - g_{\mu \nu}\left(\frac{1}{2}g^{\rho \sigma}\partial_\rho \phi \partial_\sigma \phi +V(\phi) \right).
\end{equation}
For the FRW metric, the field equations (\ref{eom}) yield the well
known Friedmann and Raychaudhuri equations,
\begin{equation}
    \centering\label{fr1}
    \frac{3}{\kappa^2} H^2 =\frac{1}{2}\dot{\phi}^2 + V(\phi),
\end{equation}
\begin{equation}
    \centering\label{fr2}
    \frac{2}{\kappa^2}\dot{H} =-\dot{\phi}^2.
\end{equation}
We assume that the slow-roll condition for the inflationary era
holds true,
\begin{equation}
    \centering\label{inf}
    V(\phi) \gg \dot{\phi}^2,
\end{equation}
thus by combining Eqs. (\ref{fr1}) and (\ref{fr2}),  we get,
\begin{equation}
    \centering\label{cond}
    \frac{|\dot{H}|}{H^2} \ll 1.
\end{equation}
The first slow-roll parameter is defined in the following way,
\begin{equation}
    \centering\label{eps1}
     \epsilon_1 = -\frac{\dot{H}}{H^2},
\end{equation}
and therefore the Friedmann equation Eq. (\ref{fr1}) becomes,
\begin{equation}
    \centering\label{h2}
    H^2 = \frac{\kappa^2}{3}V(\phi).
\end{equation}
The second slow-roll parameter is defined in the following way,
\begin{equation}
    \centering\label{eps2}
     \epsilon_2 = \frac{\ddot{\phi}}{H \dot{\phi}}.
\end{equation}
From Eq. (\ref{eomphi}) we obtain,
\begin{equation}
    \centering\label{dphi}
    \dot{\phi} = - \frac{V'}{3 H}\, ,
\end{equation}
and in addition,
\begin{equation}
    \centering\label{ddphi}
    \ddot{\phi} = - \frac{\dot{H}}{H}\dot{\phi} -V'' \frac{\dot{\phi}}{3 H}.
\end{equation}
Thus, by combining Eqs. (\ref{fr2}), (\ref{h2}), (\ref{dphi}),
(\ref{ddphi}), (\ref{eps1}) and (\ref{eps2}), we obtain the final
forms of the slow-roll parameters for the single canonical scalar
field theory,
\begin{equation}
    \centering\label{eps1 to eps}
     \epsilon_1 =  \epsilon\, ,
\end{equation}
and
\begin{equation}
    \centering\label{eps2 to eta}
     \epsilon_2 = - \eta +  \epsilon_1,
\end{equation}
where,
\begin{equation}
    \centering\label{eps}
     \epsilon = \frac{1}{2 \kappa^2}\frac{V'^2}{V^2},
\end{equation}
\begin{equation}
    \centering\label{eta}
    \eta = \frac{1}{\kappa^2}\frac{V''}{V} ,
\end{equation}
In addition, the $e$-foldings number is defined in the following
way,
\begin{equation}
    \centering\label{efold}
    N(\phi) = \int_t^{t_{end}} H d t,
\end{equation}
where $t_{end}$ denotes the end of the inflationary era, thus by
combining (\ref{h2}) and (\ref{dphi}), Eq. (\ref{efold}) takes the
form,
\begin{equation}
    \centering\label{efoldings}
    N(\phi) = \kappa^2 \int_{\phi_{f}}^{\phi_{i}} \frac{V}{V'}d \phi,
\end{equation}
with $\phi_{f}$ being the scalar field value at the end of the
inflationary regime, and $\phi_{i}$ is the value of the scalar
field at first horizon crossing. The spectral index of the
primordial scalar perturbations has the following general form
\cite{Hwang:2005hb},
\begin{equation}
    \centering\label{spec}
    n_s - 1 = -4\epsilon_1 -2\epsilon_2 ,
\end{equation}
hence by substituting (\ref{eps1 to eps}) and (\ref{eps2 to eta}),
the scalar spectral index takes the form,
\begin{equation}
    \centering\label{ns}
    n_s = 1 + 2\eta - 6\epsilon.
\end{equation}
Furthermore, the tensor-to-scalar ratio has the form,
\begin{equation}
    \centering\label{ttsr}
    r = 8 \kappa^2 \frac{\dot{\phi}^2}{H^2},
\end{equation}
which in the case at hand is equal to,
\begin{equation}
    \centering\label{r}
    r = 16 \epsilon.
\end{equation}

Now, having discussed the general structure of the scalar
inflationary theories, we consider the case that $n_s$ has a
specific functional form depending on the tensor-to-scalar ratio,
of the form,
\begin{equation}\label{generalnsr}
n_s=f(r)\, ,
\end{equation}
and $f(r)$ is a given function which can be arbitrarily chosen.
This is nothing but a second order differential equation of the
scalar field potential, of the form,
\begin{equation}\label{maindifferentialequation}
n_s = 1 + 2\eta - 6\epsilon=1+2
\frac{1}{\kappa^2}\frac{V''}{V}-6\frac{1}{2
\kappa^2}\frac{V'^2}{V^2}=f(r)=f\left(16 \frac{1}{2
\kappa^2}\frac{V'^2}{V^2}\right)\, ,
\end{equation}
which can be solved with respect to the scalar field potential.
Thus the method for obtaining desirable forms of the spectral
index is very simple. Define the desired form of the spectral
index as a function of the tensor-to-scalar ratio, and then solve
the differential equation (\ref{maindifferentialequation}) with
respect to the scalar field, to discover which potential can yield
the resulting $n_s(r)$ behavior. It is understandable that the
function $f(r)$ must be relatively simple, because the
differential equation (\ref{maindifferentialequation}) can be
highly non-trivial and non-linear for general forms of the
function $f(r)$. Thus we shall limit our analysis to cases that
can be tackled analytically.

\section{Examples of Simple $n_s(r)$ Attractors and Their Phenomenology}

In this section we shall consider four distinct attractor models
by assuming rather simple $n_s(r)$ behavior, which result to
analytically tractable phenomenology and we shall examine in
detail their inflationary phenomenology, also comparing the
results with the ACT data \cite{ACT:2025tim} and the updated
Planck/BICEP constraints \cite{BICEP:2021xfz}.

\subsection{Model I: $n_s(r)$ Attractors of the Form $n_s(r)=1-\frac{1}{3
\alpha}r^{1/2}$ and Phenomenology}

As a first example we shall consider the inflationary models which
yield the following $n_s(r)$ relation,
\begin{equation}\label{model1}
n_s(r)=1-\frac{1}{3 \alpha}r^{1/2}\, .
\end{equation}
which is the class of models discussed in the previous section
which motivated us to consider the attractor relation $n_s(r)$.
The results we expect are plateau exponential potentials, like the
Starobinsky and the Higgs model. We start from Eq. (\ref{model1}),
combining it with the differential equation
(\ref{maindifferentialequation}) and we use $f(r)=1-\frac{1}{3
\alpha}r^{1/2}$ in this case. By introducing the function
$s(\phi)$,
\begin{equation}\label{sphidefinition}
s(\phi)=\frac{V'(\phi)}{V(\phi)}\, ,
\end{equation}
and by assuming for this scenario that $s(\phi)>0$, the
differential equation (\ref{maindifferentialequation}) for
$f(r)=1-\frac{1}{3 \alpha}r^{1/2}$ can be written in this case,
\begin{equation}\label{diffmodel1}
s'+s^2-\frac{1}{2}s^2+\beta s=0\, ,
\end{equation}
where $\beta$ in this case is defined as $\beta=\frac{\sqrt{2}
\kappa }{3 \alpha }$. By solving the differential equation
(\ref{diffmodel1}) we obtain for this scenario,
\begin{equation}\label{solutionsmodel1}
s(\phi)=\frac{2\beta}{1-A\,e^{\beta \phi}}\, ,
\end{equation}
hence by recalling that $s(\phi)=\frac{V'(\phi)}{V(\phi)}$, and by
solving with respect to the scalar potential, we get,
\begin{equation}\label{potentialsolutionmodel1}
V(\phi)=V_0\frac{e^{2\beta \phi}}{\left(1-A\,e^{\beta \phi}
\right)^2}\, ,
\end{equation}
where $V_0$ and $A$ are integration constants. As usual, the
parameter $V_0$ will be constrained by the amplitude of the scalar
perturbations and does not affect the observational indices at
all.

Now let us analyze the phenomenology of the model
(\ref{potentialsolutionmodel1}) in some detail. The first
potential slow-roll index $\epsilon$ (\ref{eps}) in this case
reads,
\begin{equation}\label{epsmodel1}
 \epsilon=\frac{4}{9 \alpha ^2 \left(A e^{\frac{\sqrt{2} \kappa  \phi }{3 \alpha
 }}-1\right)^2}\, ,
\end{equation}
while the second potential slow-roll index $\eta$ (\ref{eta})
reads in this case,
\begin{equation}\label{etamodel1}
\eta=\frac{4 \left(A e^{\frac{\sqrt{2} \kappa  \phi }{3 \alpha
}}+2\right)}{9 \alpha ^2 \left(A e^{\frac{\sqrt{2} \kappa  \phi
}{3 \alpha }}-1\right)^2}\, .
\end{equation}
By solving the equation $\epsilon(\phi_f)=1$, we obtain the value
of the scalar field at the end of the inflationary regime, which
is $\phi_f=\frac{3 \alpha  \log \left(\frac{3 \alpha -2}{3 \alpha
A}\right)}{\sqrt{2} \kappa }$. Also, the value of the scalar field
at the first horizon crossing $\phi_i$ can be evaluated by using
the definition of the $e$-foldings number (\ref{efoldings}) and
also the value for $\phi_f$ we found above, so in this case we
have,
\begin{equation}\label{phiinitialmodel1}
\phi_i=\frac{3 \alpha  \log \left(\frac{9 \alpha ^2-6 \alpha -4
N}{9 \alpha ^2 A}\right)}{\sqrt{2} \kappa }\, .
\end{equation}
So in this case, the spectral index of the primordial scalar
perturbations at first horizon crossing is at leading order,
\begin{equation}\label{spectralindexmodel1}
n_s=1-\frac{2}{N}+\frac{3 \alpha }{N^2}\, ,
\end{equation}
while the corresponding tensor-to-scalar ratio reads,
\begin{equation}\label{tensortoscalarmodel1}
r=\frac{36 \alpha ^2}{N^2}\, .
\end{equation}
For $N=60$, the maximum value allowed for $\alpha$ in order for
the tensor-to-scalar ratio to be compatible with the Planck/BICEP
constraint is $\alpha\simeq 1.991$. So for this value we have
$n_s\simeq 0.968247$ and $r\simeq 0.0359708$. Therefore, this
class of models can never be compatible with the ACT data,
something which we expected, since for the plateau potentials we
always have $n_s=1-\frac{2}{N}$ at leading order, as it can also
be seen by looking at Eq. (\ref{spectralindexmodel1}). Before
closing, let us also evaluate the effective equation of state
parameter (EoS), defined as,
\begin{equation}\label{EoS}
w_{eff}=-1-\frac{2}{3}\frac{\dot{H}}{H^2}=-1+\frac{r}{24}\, ,
\end{equation}
which we expect it to be $w_{eff}=-1/3$ at the end of inflation,
so at $N=0$ or equivalently at $\phi=\phi_f$, while it must be
near its de Sitter value $w_{eff}=-1$ at the beginning of
inflation, so at $\phi=\phi_i$, or at $N\sim 60$. In this case at
$\phi=\phi_i$, for $\alpha\simeq 1.991$ we have
$w_{eff}=-0.998501$, while at $\phi=\phi_f$ we have
$w_{eff}=-0.3333$ which is the expected behavior.

\subsection{Model II: $n_s(r)$ Attractors of the Form $n_s(r)=1-\alpha r$ and Phenomenology}

\begin{figure}
\centering
\includegraphics[width=28pc]{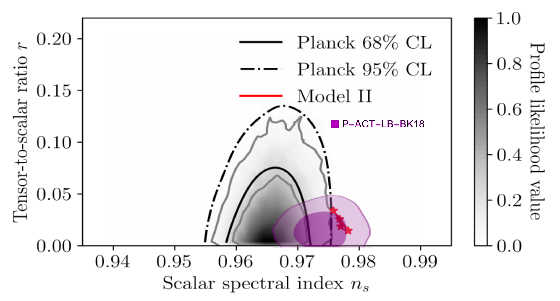}
\caption{Marginalized curves of the Planck 2018 data and the model
(\ref{potentialsolutionmodel12}), confronted with the ACT data,
the Planck 2018 data, and the updated Planck constraints on the
tensor-to-scalar ratio for $N=50$ and $\alpha$ in the range
$\alpha=[0.68,1.5]$. }\label{plot1}
\end{figure}

Now we consider the inflationary models which yield the following
$n_s(r)$ relation,
\begin{equation}\label{model12}
n_s(r)=1-\alpha\,r\, .
\end{equation}
Let us see which class of potentials lead to the $n_s(r)$ relation
of Eq. (\ref{model12}). We start from Eq. (\ref{model12}),
combining it with the differential equation
(\ref{maindifferentialequation}) and we use in this case
$f(r)=1-\alpha\,r$. By using the function $s(\phi)$ of Eq.
(\ref{sphidefinition}), the differential equation
(\ref{maindifferentialequation}) for $f(r)=1-\alpha r$ can be
written in this case,
\begin{equation}\label{diffmodel12}
s'-\frac{8\alpha-1}{2}s^2=0\, .
\end{equation}
By solving the differential equation (\ref{diffmodel12}) we obtain
for this scenario,
\begin{equation}\label{solutionsmodel12}
s(\phi)=\frac{1}{A\phi+c_1}\, ,
\end{equation}
where $A=\frac{8\alpha-1}{2}$ and $c_1$ is an integration
constant. Hence by recalling that
$s(\phi)=\frac{V'(\phi)}{V(\phi)}$, and by solving with respect to
the scalar potential, we get in this case,
\begin{equation}\label{potentialsolutionmodel12}
V(\phi)=V_0\left(A \phi+c_1\right)^{1/A}\, ,
\end{equation}
where $V_0$ and $A$ are integration constants. Note that if
$\alpha=1/8$, hence $A=0$ the differential equation
(\ref{diffmodel12}) would yield the solution $V(\phi)=V_0\,e^{S_0
\phi}$ which is a simple exponential inflation, leading to
non-viable inflationary phenomenology and to eternal inflation.
Now let us analyze the phenomenology of the model
(\ref{potentialsolutionmodel12}) in some detail. The first
potential slow-roll index $\epsilon$ (\ref{eps}) in this case
reads,
\begin{equation}\label{epsmodel12}
 \epsilon=\frac{2}{\kappa ^2 ((8 \alpha -1) \phi +2 c_1)^2}\, ,
\end{equation}
while the second potential slow-roll index $\eta$ (\ref{eta})
reads in this case,
\begin{equation}\label{etamodel12}
\eta=\frac{6-16 \alpha }{\kappa ^2 ((8 \alpha -1) \phi +2
c_1)^2}\, .
\end{equation}
By solving the equation $\epsilon(\phi_f)=1$, we obtain the value
of the scalar field at the end of the inflationary regime, which
is,
\begin{equation}\label{phifmodel12}
\phi_f=\frac{\sqrt{2} \sqrt{64 \alpha ^2 \kappa ^2-16 \alpha
\kappa ^2+\kappa ^2}-16 \alpha  c_1 \kappa ^2+2 c_1 \kappa ^2}{64
\alpha ^2 \kappa ^2-16 \alpha  \kappa ^2+\kappa ^2}\, .
\end{equation}
Also, the value of the scalar field at the first horizon crossing
$\phi_i$ can be evaluated by using the definition of the
$e$-foldings number (\ref{efoldings}) and also for the value of
$\phi_f$ we found above, so in this case we have,
\begin{equation}\label{phiinitialmodel12}
\phi_i=\frac{2 \left(\kappa  \sqrt{8 \alpha  \Gamma -\Gamma +c_1^2
\kappa ^2+8 \alpha  N-N}-c_1 \kappa ^2\right)}{8 \alpha \kappa
^2-\kappa ^2}\, ,
\end{equation}
where $\Gamma$ is defined as follows,
\begin{equation}\label{gammaextra}
\Gamma=\kappa ^2 \left(2 \alpha  \phi_f^2+c_1
\phi_f-\frac{\phi_f^2}{4}\right)\, .
\end{equation}
So in this case, the spectral index of the primordial scalar
perturbations at first horizon crossing is at leading order,
\begin{equation}\label{spectralindexmodel12}
n_s=1-\frac{8 \alpha }{(8 \alpha -1) N}+\frac{4 \alpha }{(8 \alpha
-1)^2 N^2}\, ,
\end{equation}
while the corresponding tensor-to-scalar ratio reads,
\begin{equation}\label{tensortoscalarmodel12}
r=\frac{8}{(8 \alpha -1) N}-\frac{4}{(8 \alpha -1)^2 N^2}\, .
\end{equation}
Now this model can be compatible with both the ACT constraint on
the scalar spectral index and the updated BICEP/Planck constraint.
For example if we choose, $N=50$ and $\alpha=0.715$ we get
$n_s=0.975814$ and $r=0.0338266$, well fitted in the ACT and
BICEP/Planck data. We can have a better idea of the viability of
this model by looking in Fig. \ref{plot1} where we present the
confrontation of the model (\ref{potentialsolutionmodel12}) with
the Planck 2018 data and the updated ACT data, choosing $\alpha$
in the range $\alpha=[0.68,1.5]$ and $N=50$. Note that for
$\alpha<0.68$ the BICEP/Planck constraint is violated. So for
$\alpha=[0.68,1.5]$, as it can be seen in Fig. \ref{plot1} the
model (\ref{potentialsolutionmodel12}) is well fitted within the
ACT and the BICEP/Planck data.
Before closing, let us also evaluate the EoS parameter of Eq.
(\ref{EoS}) at the end and at the beginning of inflation. In this
case at $\phi=\phi_i$, for $\alpha\simeq 0.715$ we have
$w_{eff}=-0.99859$, while at $\phi=\phi_f$ we have
$w_{eff}=-0.3333$ which is the expected behavior.

\subsection{Model III: $n_s(r)$ Attractors of the Form $n_s(r)=1-\beta  r-\gamma  \sqrt{r}$ and Phenomenology: Scenario I $s(\phi)<0$}

\begin{figure}
\centering
\includegraphics[width=28pc]{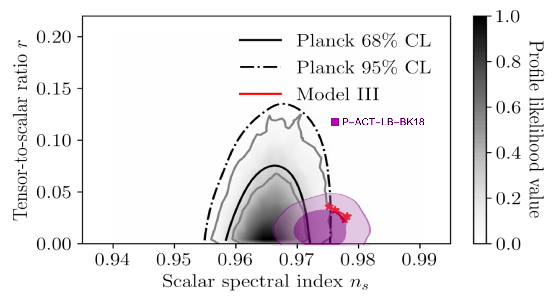}
\caption{Marginalized curves of the Planck 2018 data and the model
(\ref{potentialsolutionmodel123}), confronted with the ACT data,
the Planck 2018 data, and the updated Planck constraints on the
tensor-to-scalar ratio for $N=50$, $K=10$, $\gamma=-0.0000001$ and
$\beta$ in the range $\beta=[0.68,0.98]$.}\label{plot2}
\end{figure}

Now we consider the inflationary models which yield the following
$n_s(r)$ relation,
\begin{equation}\label{model123}
n_s(r)=1-\beta  r-\gamma  \sqrt{r}\, .
\end{equation}
Let us see which class of potentials lead to the $n_s(r)$ relation
of Eq. (\ref{model123}). We start from Eq. (\ref{model123}),
combining it with the differential equation
(\ref{maindifferentialequation}) and we use in this case
$f(r)=1-\beta  r-\gamma  \sqrt{r}$. By using the function
$s(\phi)$ of Eq. (\ref{sphidefinition}), the differential equation
(\ref{maindifferentialequation}) for $f(r)=1-\beta  r-\gamma
\sqrt{r}$ can be written in this case,
\begin{equation}\label{diffmodel123}
s'=A\,s^2+B\,s\, ,
\end{equation}
where $A=\frac{1-8\beta}{2}$ and $B=\frac{\gamma \kappa
\sqrt{8}}{2}$. By solving the differential equation
(\ref{diffmodel123}) and by solving with respect to the scalar
potential, for $s(\phi)<0$ in this case, we get,
\begin{equation}\label{potentialsolutionmodel123}
V(\phi)=V_0\left(1-A K e^{B\phi}\right)^{-1/A}\, ,
\end{equation}
where $V_0$, $K$ are integration constants. In this case, it is
vital to have $s(\phi)<0$ and we must check in the end for the
values for which the viability comes, whether $s(\phi)<0$ for
$\phi=\phi_i$ and $\phi=\phi_f$. Of course it must hold true for
all the values of $N$ in the range $N=[0,60]$. Now let us analyze
the phenomenology of the model (\ref{potentialsolutionmodel123})
in some detail. The first potential slow-roll index $\epsilon$
(\ref{eps}) in this case reads,
\begin{equation}\label{epsmodel123}
 \epsilon=\frac{4 \gamma ^2 K^2 e^{2 \sqrt{2} \gamma  \kappa  \phi }}{(1-8 \beta )^2 \left(K e^{\sqrt{2} \gamma  \kappa  \phi }-1\right)^2}\, ,
\end{equation}
while the second potential slow-roll index $\eta$ (\ref{eta})
reads in this case,
\begin{equation}\label{etamodel123}
\eta=\frac{4 \gamma ^2 K e^{\sqrt{2} \gamma  \kappa  \phi }
\left(-8 \beta +2 K e^{\sqrt{2} \gamma  \kappa  \phi
}+1\right)}{(1-8 \beta )^2 \left(K e^{\sqrt{2} \gamma  \kappa
\phi }-1\right)^2}\, .
\end{equation}
By solving the equation $\epsilon(\phi_f)=1$, we obtain the value
of the scalar field at the end of the inflationary regime, which
is,
\begin{equation}\label{phifmodel123}
\phi_f=\frac{\log \left(\frac{8 \beta -1}{K (8 \beta +2 \gamma
-1)}\right)}{\sqrt{2} \gamma  \kappa }\, .
\end{equation}
Also, the value of the scalar field at the first horizon crossing
$\phi_i$ can be evaluated by using the definition of the
$e$-foldings number (\ref{efoldings}) and also for the value of
$\phi_f$ we found above, so in this case we have,
\begin{equation}\label{phiinitialmodel123}
\phi_i=\frac{8 \beta  W\left(\frac{e^{-\frac{4 \gamma ^2 (\Gamma
+N)}{8 \beta -1}}}{K}\right)-W\left(-\frac{e^{-\frac{4 \gamma ^2
(\Gamma +N)}{8 \beta -1}}}{K}\right)+4 \gamma ^2 (\Gamma
+N)}{\sqrt{2} (8 \beta -1) \gamma  \kappa }\, ,
\end{equation}
where $\Gamma$ is defined as follows,
\begin{equation}\label{gammaextra3}
\Gamma=\frac{(8 \beta -1) \kappa  \left(\frac{e^{-\sqrt{2} \gamma
\kappa  \phi_f}}{\sqrt{2} \gamma  \kappa }+K \phi_f\right)}{2
\sqrt{2} \gamma  K}\, ,
\end{equation}
and also $W(z)$ in Eq. (\ref{phiinitialmodel123}) is the Lambert
function. So in this case, we can easily obtain the scalar
spectral index and the tensor-to-scalar ratio in closed analytic
form, but these are too extended to quote here. Now this model can
be compatible with both the ACT constraint on the scalar spectral
index and the updated BICEP/Planck constraint. For example if we
choose, $N=50$, $K=10$, $\gamma=-0.0000001$ and $\beta=0.74$ we
get $n_s=0.97618$ and $r=0.032471$, well fitted in the ACT and
BICEP/Planck data. We can have a better idea of the viability of
this model too by looking in Fig. \ref{plot2} where we present the
confrontation of the model (\ref{potentialsolutionmodel123}) with
the Planck 2018 data and the updated ACT data, for $N=50$, $K=10$,
$\gamma=-0.0000001$ and $\beta$ in the range $\beta=[0.68,0.98]$.
Note that if $\beta<0.68$ the BICEP/Planck constraint is violated.
So for $N=50$, $K=10$, $\gamma=-0.0000001$ and $\beta$ in the
range $\beta=[0.68,0.98]$, as it can be seen in Fig. \ref{plot2}
the model (\ref{potentialsolutionmodel123}) is well fitted within
the ACT and the BICEP/Planck data.
Before closing, let us also evaluate the EoS parameter of Eq.
(\ref{EoS}) at the end and at the beginning of inflation. In this
case at $\phi=\phi_i$, for $N=50$, $K=10$, $\gamma=-0.0000001$ and
$\beta=0.74$ we have $w_{eff}=-0.99864$, while at $\phi=\phi_f$ we
have $w_{eff}=-0.3333$ which is again the expected behavior. Also
it can easily be checked that the constraint $s(\phi)<0$ is well
respected for all the values of the scalar field in the range
$\phi=[\phi_i,\phi_f]$. In fact at $\phi=\phi_i$ we have
$s(\phi_i)=-0.0637095 \kappa$ and at $\phi=\phi_k$ we have
$s(\phi_f)=-\sqrt{2} \kappa$.

\subsection{Model IV: $n_s(r)$ Attractors of the Form $n_s(r)=1-\beta  r-\gamma  \sqrt{r}$ and Phenomenology: Scenario II $s(\phi)>0$}

\begin{figure}
\centering
\includegraphics[width=28pc]{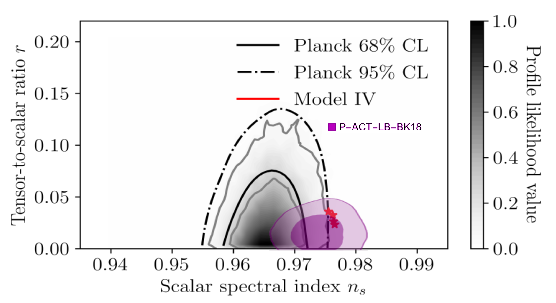}
\caption{Marginalized curves of the Planck 2018 data and the model
(\ref{potentialsolutionmodel1234}), confronted with the ACT data,
the Planck 2018 data, and the updated Planck constraints on the
tensor-to-scalar ratio for $N=50$, $K=-10$, $\gamma=-0.0000001$
and $\beta$ in the range $\beta=[0.68,0.98]$.}\label{plot3}
\end{figure}

Now we consider the same inflationary models which yield the
$n_s(r)$ relation of Eq. (\ref{model123}), namely,
\begin{equation}\label{model1234}
n_s(r)=1-\beta  r-\gamma  \sqrt{r}\, ,
\end{equation}
but in this case we assume $s(\phi)>0$ during the duration of
inflation. In this case, the differential equation
(\ref{maindifferentialequation}) becomes,
\begin{equation}\label{diffmodel1234}
s'=A\,s^2-B\,s\, ,
\end{equation}
where $A=\frac{1-8\beta}{2}$ and $B=\sqrt{2}\gamma \kappa$. By
solving the differential equation (\ref{diffmodel1234}) and by
solving with respect to the scalar potential, for $s(\phi)>0$ in
this case, we get,
\begin{equation}\label{potentialsolutionmodel1234}
V(\phi)=V_0\frac{e^{\frac{B\phi}{A}}}{\left(A-K
e^{B\phi}\right)^{1/A}}\, ,
\end{equation}
where $V_0$, $K$ are integration constants. In this case, it is
vital to have $s(\phi)>0$ and we also must check this in the end
for the values for which the viability comes, whether $s(\phi)>0$
for $\phi=\phi_i$ and $\phi=\phi_f$. Of course, as in the previous
case, it must hold true for all the values of $N$ in the range
$N=[0,60]$. Now let us again analyze the phenomenology of the
model (\ref{potentialsolutionmodel1234}) in some detail. The first
potential slow-roll index $\epsilon$ (\ref{eps}) in this case
reads,
\begin{equation}\label{epsmodel1234}
 \epsilon=\frac{4 \gamma ^2}{\left(8 \beta +2 K e^{\sqrt{2} \gamma  \kappa  \phi }-1\right)^2}\, ,
\end{equation}
while the second potential slow-roll index $\eta$ (\ref{eta})
reads in this case,
\begin{equation}\label{etamodel1234}
\eta=\frac{8 \gamma ^2 \left(K e^{\sqrt{2} \gamma  \kappa  \phi
}+1\right)}{\left(8 \beta +2 K e^{\sqrt{2} \gamma  \kappa  \phi
}-1\right)^2}\, .
\end{equation}
By solving the equation $\epsilon(\phi_f)=1$, we obtain the value
of the scalar field at the end of the inflationary regime, which
is,
\begin{equation}\label{phifmodel1234}
\phi_f=\frac{\log \left(\frac{-8 \beta -2 \gamma +1}{2
K}\right)}{\sqrt{2} \gamma  \kappa }\, .
\end{equation}
Also, the value of the scalar field at the first horizon crossing
$\phi_i$ in this case is,
\begin{equation}\label{phiinitialmodel1234}
\phi_i=\frac{(1-8 \beta ) W\left(\frac{2 K e^{-\frac{4 \gamma ^2
(\Gamma +N)}{8 \beta -1}}}{8 \beta -1}\right)-4 \gamma ^2 (\Gamma
+N)}{\sqrt{2} (8 \beta -1) \gamma  \kappa }\, ,
\end{equation}
where $\Gamma$ is defined as follows,
\begin{equation}\label{gammaextra4}
\Gamma=-\frac{\kappa  \left(8 \beta  \phi_f+\frac{\sqrt{2} K
e^{\sqrt{2} \gamma  \kappa  \phi_f}}{\gamma  \kappa
}-\phi_f\right)}{2 \sqrt{2} \gamma }\, ,
\end{equation}
and $W(z)$ in Eq. (\ref{phiinitialmodel1234}) is again the Lambert
function. In this case too we can obtain the scalar spectral index
and the tensor-to-scalar ratio in closed analytic form, but these
are again too extended to quote here. Now this model too can be
compatible with both the ACT constraint on the scalar spectral
index and the updated BICEP/Planck constraint. For example if we
choose, $N=50$, $K=-10$, $\gamma=-0.0000001$ and $\beta=0.74$ we
get $n_s=0.97624$ and $r=0.032471$, well fitted in the ACT and
BICEP/Planck data, and similar to the ones of the model with
$s(\phi)<0$. We can have a better idea of the viability of this
model too by looking in Fig. \ref{plot3} where we present the
confrontation of the model (\ref{potentialsolutionmodel1234}) with
the Planck 2018 data and the updated ACT data, for $N=50$,
$K=-10$, $\gamma=-0.0000001$ and $\beta$ in the range
$\beta=[0.68,0.98]$. Note that in this case too, if $\beta<0.68$
the BICEP/Planck constraint is violated. So for $N=50$, $K=-10$,
$\gamma=-0.0000001$ and $\beta$ in the range $\beta=[0.68,0.98]$,
as it can be seen in Fig. \ref{plot3} the model
(\ref{potentialsolutionmodel1234}) is well fitted within the ACT
and the BICEP/Planck data.
Let us also evaluate the EoS parameter of Eq. (\ref{EoS}) at the
end and at the beginning of inflation. In this case at
$\phi=\phi_i$, for $N=50$, $K=-10$, $\gamma=-0.0000001$ and
$\beta=0.74$ we have $w_{eff}=-0.99865$, while at $\phi=\phi_f$ we
have $w_{eff}=-0.3333$ which is again the expected behavior and
similar to the previous model. Also it can easily be checked that
the constraint $s(\phi)>0$ is well respected for all the values of
the scalar field in the range $\phi=[\phi_i,\phi_f]$. In fact at
$\phi=\phi_i$ we have $s(\phi_i)=0.0637095 \kappa$ and at
$\phi=\phi_k$ we have $s(\phi_f)=\sqrt{2} \kappa$.

\section{Conclusions}

In this article we sought for general classes of single scalar
field inflationary models which can be compatible with the ACT and
Planck/BICEP constraints using a top-bottom approach. With the
top-bottom approach we mean that we did not start studying the
phenomenology of inflationary models from the scalar potential
itself, instead we started by defining the functional dependence
of the spectral index of the scalar perturbations as a function of
the tensor-to-scalar ratio, that is the function $n_s(r)=f(r)$.
The relation $n_s(r)=f(r)$ is a differential equation for the
single scalar field theory and thus by providing the function
$f(r)$, it is possible that the potential is found for some
choices of the function $f(r)$. We presented three classes of
models which can be viable with the ACT and BICEP/Planck data. Our
original motivation was inspired by the plateau potentials which
result to a universal relation shared for all the potentials that
behave as plateau potentials asymptotically. These models include
the Higgs and Starobinsky inflation as well as the $a$-attractor
models and in these models one has $n_s(r)=1-\alpha r^{1/2}$. The
classes of $n_s(r)$ attractor models which we considered have a
relatively simple $n_s(r)$ relation, which facilitates finding the
solution of the differential equation in closed form. The classes
of models we considered are variants of the form $n_s(r)=\gamma\pm
\beta r \pm \alpha r^{1/2}$ and the resulting phenomenologies were
found to be viable and compatible with the ACT data. It would be
useful to investigate whether these models can be a part of some
no-scale supergravity theory, but this task stretches beyond the
aims and scopes of this introductory article.

\end{document}